\documentclass[aps,nofootinbib,twocolumn,superscriptaddress]{revtex4-2}

\usepackage[utf8]{inputenc}
\usepackage[english]{babel}
\usepackage{amssymb,amsmath,amsfonts,amsthm,mathtools,dsfont}
\usepackage{hyperref}
\usepackage{physics, xparse}
\newcommand{\rme}{\mathrm e}

\newcommand{\rmi}{\mathrm i}

\usepackage{graphicx}
\usepackage{float}
\usepackage{enumitem}
\usepackage{xcolor}
\usepackage{appendix}
\usepackage{bbm}
\usepackage{bbold}
\usepackage{xcolor}
\usepackage{comment}

\newcommand{\beq}{\begin{equation}}
\newcommand{\eeq}{\end{equation}}
\newcommand{\bal}{\begin{aligned}}
\newcommand{\eal}{\end{aligned}}

\usepackage{tikz}
\usetikzlibrary{shapes.geometric}

\newcommand{\be}{\begin{equation}}
\newcommand{\ee}{\end{equation}}
\newcommand{\ben}{\begin{equation*}}
\newcommand{\een}{\end{equation*}}
\newcommand{\ba}{\begin{aligned}}
\newcommand{\ea}{\end{aligned}}

\renewcommand{\AA}{\mathcal{A}}

\newcommand{\bne}{\begin{equation}}
\newcommand{\ene}{\end{equation}}

\newcommand{\cC}{\mathcal{C}}

\definecolor{darkpastelgreen}{rgb}{0.01, 0.75, 0.24}

 \hypersetup{
colorlinks=true
  ,urlcolor=blue
  ,anchorcolor=blue
  ,citecolor=blue
  ,filecolor=blue
  ,linkcolor=blue
  ,menucolor=blue
  ,linktocpage=true
  ,pdfproducer=medialabs
}
\begin{document}
\title{
Krylov complexity is not a measure of distance between states or operators
}

\author{Sergio E. Aguilar-Gutierrez}
\email{sergio.ernesto.aguilar@gmail.com}
\affiliation{Institute for Theoretical Physics, KU Leuven, 3001 Leuven, Belgium}

\author{Andrew Rolph}
\email{andrew.d.rolph@gmail.com}
\affiliation{Institute for Theoretical Physics, University of Amsterdam, Science Park 904, 1090 GL Amsterdam, The Netherlands}
\affiliation{Theoretische Natuurkunde, Vrije Universiteit Brussel (VUB) and The International Solvay Institutes, Pleinlaan 2, B-1050 Brussels, Belgium}

\begin{abstract}
We ask whether Krylov complexity is mutually compatible with the circuit and Nielsen definitions of complexity.
We show that the Krylov complexities between three states fail to satisfy the triangle inequality and so cannot be a measure of distance: there is no possible metric for which Krylov complexity is the length of the shortest path to the target state or operator. We show this explicitly in the simplest example, a single qubit, and in general.
\end{abstract}

\maketitle

\section{Introduction}

Intuitively, complexity is a measure of how `hard' it is to prepare a state or operator. Attempts to make intuition precise have led to several different definitions of complexity. 
Can these definitions be unified?

One definition is circuit complexity: the minimum number of elementary gates from a universal gate set needed to prepare a given unitary operator, up to some tolerance. This definition has shortcomings: it depends on the choice of gate set, it does not apply to continuum theories, and the minimisation step is often impossible in practice. 

Nielsen complexity is a geometrisation of circuit complexity and is defined as the length of the shortest path to the target unitary operator \cite{Nielsen2005,nielsen2006quantum,dowling2008geometry}. 
Distances between points are specified by a cost function
and the metric that comes from the cost function is a Finsler metric, which is a generalisation of Riemannian metrics to non-quadratic forms. 
Nielsen complexity is still difficult to apply to continuum theories and its definition is still ambiguous as it depends on the choice of metric.

A third definition is Krylov complexity. The Krylov complexity of a time-evolved pure state $\ket{\psi(t)}$ is defined as
\begin{equation}
    \cC(t)=\sum_n n\abs{\bra{\psi(t)}\ket{K_n}}^2~,
\end{equation}
where $\ket{K_n}$ is the orthonormal, ordered Krylov basis. First established as a measure of the complexity of time-evolved operators in \cite{Parker:2018yvk} and states in \cite{Balasubramanian:2022tpr}, it describes the average position of the time-evolved state or operator along a 1D chain of Krylov basis states or operators. A significant advantage of Krylov complexity over other definitions is that it is unambiguously defined, and it has already found numerous applications~\cite{Barbon:2019wsy, Rabinovici:2020ryf, Rabinovici:2021qqt, Rabinovici:2022beu,Erdmenger:2023wjg}.

In this paper, we ask whether Krylov complexity is compatible with the circuit and Nielsen definitions of complexity. This is important because if we show that, with a suitable choice of metric, Krylov complexity equals Nielsen complexity, then we have a unification of definitions. If they are not compatible, then Krylov complexity is a fundamentally different object than previous definitions of complexity that \textit{are} measures of distance to the target state or operator.

What would be the Nielsen geometry for Krylov complexity?
For a given Hamiltonian, is there a metric we can place on a Hilbert space such that Krylov complexity equals the length of the shortest path between two states related by time evolution?
When calculating Nielsen complexity, one specifies a metric on the Hilbert space and calculates minimal length geodesics between fixed pairs of points. Our task is essentially the inverse: determine the metric, assuming that Krylov complexity, which is defined for any pair of states related by time evolution, equals the length of the shortest path between the pair.   

The connection between geometry and Krylov complexity was previously investigated in~\cite{Caputa:2021ori}, but we will take the opposite approach. The Fubini-Study metric was assumed
and it was shown, for some specific cases, that Krylov complexity equals a volume of Hilbert space, rather than a length. Nonetheless, this does not immediately rule out the possibility that there is a different metric for which Krylov complexity equals Nielsen complexity. While our results are not inconsistent with those from~\cite{Caputa:2021ori}, we do not assume a particular metric, instead, we ask if there is any metric for which Krylov complexity equals the length of a minimal-length geodesic.

Krylov complexity has also been identified with the dilaton in JT gravity~\cite{Chattopadhyay:2023fob}, the relative height for SU(2) symmetric systems with a planar metric on a deformed sphere \cite{Lv:2023jbv}, and with wormhole length~\cite{Rabinovici:2023yex}. These results also do not rule out the possibility that Krylov complexity equals distance in a Nielsen geometry.

We will begin by reviewing the definition of Krylov complexity. We will then show that Krylov complexity fails to satisfy the triangle inequality
\begin{equation}
    \label{eq: s triangle}\cC(t_A, t_B ) + \cC(t_B, t_C) \geq \cC(t_A , t_C)~.
\end{equation}
where $\cC(t_A,t_B)$ is defined to be the K-complexity between two states related by time-evolution: $\rme^{-iHt_A}\ket{\psi_0}$ and target state $\rme^{-iHt_B}\ket{\psi_0}$, with $\ket{\psi_0}$ a reference state. The triangle inequality is a fundamental property of distance: the shortest path from $A$ to $C$ can only increase in length if one adds an intermediate stop $B$. 
Since Krylov complexity is only defined for states related by time evolution, to be clear, with the triangle inequality~\eqref{eq: s triangle}, we are comparing Krylov complexities between three states in the same one-parameter family of states related by time-evolution, and testing whether they satisfy a defining property of distance. 

We explicitly show that~\eqref{eq: s triangle} is not satisfied in the simplest non-trivial example, the single qubit, and for general higher dimensional Hilbert spaces. For an intermediate step, we prove that Krylov complexity is time-translation invariant, i.e. that the complexity between two states related by time-evolution is only a function of the time separation:
\begin{equation}\label{eq:equal K complexity}
    \cC\qty(t_A,~t_B)=\cC\qty(t_B-t_A)~,
\end{equation}
and so $\cC(t)$, which is shorthand for $\cC(0,t)$, must be a subadditive function of time to satisfy the triangle inequality~\eqref{eq: s triangle}. We show that Krylov complexity is not generally a subadditive function of time, by perturbatively solving its time evolution equation for small time intervals. 

Both circuit and Nielsen complexity satisfy the triangle inequality. The violation of the triangle inequality by Krylov complexity implies that it cannot be a measure of distance in any metric space. Krylov complexity and Nielsen complexity are not mutually compatible.

\section{Review of Krylov complexity}
We review the definition of Krylov state complexity introduced in \cite{Balasubramanian:2022tpr}.
Consider the time evolution of a pure initial state $\ket{\psi(t)}=\rme^{-\rmi Ht}\ket{\psi_0}$. Defined by the initial state and the Hamiltonian, the Krylov basis $\ket{K_n}$ is an ordered, orthonormal set of states in the Hilbert space. The zeroth state in this basis is the initial state
$\ket{K_0}:=\ket{\psi_0}$
and $\ket{K_{n}}$ for $n \geq 1$ are defined recursively through the Lanczos algorithm, with orthonormality ensured, by construction, through a Gram–Schmidt process:
\begin{align}
    \ket{A_{n+1}}&:= (H - a_n)\ket{K_n} - b_n \ket{K_{n-1}}~, \label{eq:lanczos}\\
    \ket{K_n} &:= b_n^{-1}\ket{A_n}
\end{align}
where $a_n$ and $b_n$ are the Lanczos coefficients:
\begin{equation}
    a_{n} := \bra{K_n} H \ket{K_n}, \qquad b_{n} := (\braket{A_n})^{1/2}~,
\end{equation}
The Krylov basis $\ket{K_n}$ is special because it minimises $\sum_n c_n\abs{\bra{\psi(t)}\ket{B_n}}^2$, with c$_n$ any monotonically increasing real sequence, amongst all orthonormal bases $\ket{B_n}$ \cite{Balasubramanian:2022tpr}. 

We can express $\ket{\psi(t)}$ in the Krylov basis:
\begin{equation}
    \ket{\psi(t)} = \sum^{\mathcal{K}}_{n=0} \psi_n (t) \ket{K_n}~,
\end{equation}
where $\mathcal{K}$ is the Krylov space dimension, which is smaller than or equal to the Hilbert space dimension. The Hamiltonian is tridiagonal in the Krylov basis, so the wavefunctions obey a simple Schr\"odinger equation:
\begin{equation} \label{eq:Schro}
    \rmi\partial_t \psi_n(t) = a_n \psi_n (t) + b_{n+1}\psi_{n+1}(t) + b_n \psi_{n-1}(t)~.
\end{equation}
Krylov complexity is defined as
\begin{equation} \label{eq:Complexity}
   \cC(t) := \sum_n n p_n (t) ~,
\end{equation}
where $p_n$ is a probability distribution
\begin{equation}\label{eq:prob def}
    p_n(t) := |\psi_n(t)|^2,\quad \sum_np_n(t)=1~.
\end{equation}
Why is~\eqref{eq:Complexity} a measure of complexity? Roughly, the complexity of each basis state $\ket{K_n}$ increases with $n$ because it involves more applications of the Hamiltonian, $\ket{K_n} \approx H^n \ket{\psi(0)}$, so Krylov complexity quantifies the average position along a 1D chain of increasingly complex basis states. 

While Krylov complexity is the average position of a particle along a 1D chain, which sounds like a measure of distance,
to be a distance in the mathematical sense it must satisfy the triangle inequality. 

In the triangle inequality~\eqref{eq: s triangle}, we used the notation $C(t_A,t_B)$ for the Krylov complexity between two states constructed from the same reference state evolved for different amounts of time. In this notation, $\cC(t)$ as defined in~\eqref{eq:Complexity} is shorthand for $\cC(0,t)$, and to calculate $\cC(t_A, t_B)$, since we have a different initial state, we have to find the new Krylov basis starting from $\ket{K_0} = e^{-iHt_A} \ket{\psi_0}$.

Krylov operator complexity is defined in a similar way as to state complexity, with the replacements (see~\cite{Parker:2018yvk} for more details)
\bne H\xrightarrow{}\mathcal{L},\quad \ket{K_n}\rightarrow |O_n),\quad a_n\rightarrow 0~ \ene
The results we will show for Krylov state complexity carry over to operator complexity because our derivation does not depend on whether the Krylov basis vectors are states or operators.
\section{Time-translation invariance}
Next, we show that Krylov complexity is time-translation invariant. This is a useful property to show because the triangle inequality involves Krylov complexities with initial states that are different but related by time evolution. The time-evolution of a reference state generates a curve through projective Hilbert space, and by time-translation invariance we mean that the Krylov complexity between any pair of points on that curve is a function of the time separation:\footnote{To reiterate, $\cC (0,t'-t)$, as we define it after~\eqref{eq: s triangle}, is equivalent to the more commonly used $\cC (t'-t)$.}
\bne \label{eq:TimeTranslation} \cC\qty(t,~t')=\cC\qty(t'-t). \ene

To see this, let us compare the Krylov bases built from two initial states that are related by time evolution. It is straightforward to show, following the the Lanczos algorithm (\ref{eq:lanczos}), that, if $\ket{K'_0}=\rme^{-\rmi Ht}\ket{K_0}$, then
\begin{equation}
\ket{K_n'}=\rme^{-\rmi Ht}\ket{K_n}~
\end{equation}
for all $n$, and the Lanczos coefficients are the same for these two initial states.
Next, we use the definition of the Krylov basis to derive a relation between the amplitudes: 
\begin{equation}
    \begin{aligned}
        \psi_n'(t'):&=\bra{K'_n}\ket{\psi(t')}=\bra{K_n}\ket{\psi(t'-t)}=\psi_n(t'-t)~.
    \end{aligned}
\end{equation} 
This means
\begin{equation}
    \begin{aligned}
    \cC\qty(t,~t')&=\sum_nn\abs{\psi'_n(t')}^2=\sum_nn\abs{\psi_n(t'-t)}^2~,
\end{aligned}
\end{equation}
which then proves \eqref{eq:TimeTranslation}. A consequence of time-translation invariance is that the triangle inequality~\eqref{eq: s triangle} reduces to 
the requirement that Krylov complexity is 
a subadditive function of time:
\bne \cC (t_B - t_A) + \cC (t_C - t_B) \geq \cC (t_C - t_A). \ene

\section{The Bloch sphere}
The simplest setup to consider is a single qubit with a two-dimensional Hilbert space. We will determine the Krylov complexity for a time-evolved state, starting from an arbitrary initial position on the Bloch sphere. Then, we will try to determine whether there is a metric on the sphere that would give the K-complexity as the length of the shortest path between two points on the sphere.

There is a natural metric on projective Hilbert spaces like the Bloch sphere, the Fubini-Study (FS) metric \cite{fubini1904sulle,study1905kurzeste}, which is the unique, up-to-rescaling, homogeneous metric. However, in the context of Nielsen complexity, it is not necessary to assume this metric and our approach is to look for any metric for which Krylov complexity equals Nielsen complexity on the Bloch sphere.

Take the initial state to be an arbitrary pure state:
\bne \ket{\psi_0} = \cos (\theta_0 /2) \ket{E_0} + e^{i\phi_0} \sin (\theta_0 /2) \ket{E_1} \ene
The angles denote the position on the Bloch sphere: $0 \leq \theta_0 \leq \pi$ and $0 \leq \phi_0 \leq 2 \pi$. The Krylov basis is
$\ket{K_0} = \ket{\psi_0}$,
and $\ket{K_1}$ is the state that is orthonormal to this. 

\begin{figure}
    \centering
    \includegraphics[width=0.3\textwidth]{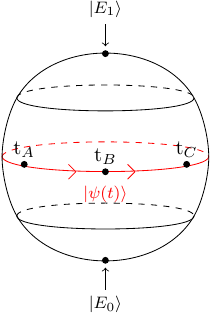}
    \caption{
    The time evolution of a pure state on the Bloch sphere. 
    The poles of the sphere are the energy eigenstates $\ket{E_0}$ and $\ket{E_1}$. The time evolution of an arbitrary pure state generates the red curves. By definition, Krylov complexity gives us the complexity between states related by time evolution, on the same red curve, and we show that it is inconsistent with a measure of distance because it fails to satisfy the triangle inequality (\ref{eq: s triangle}).
    }
    \label{fig:Bloch sphere}
\end{figure}

The position of the time evolved state 
\bne \ket{\psi(t)} = e^{-iHt}\ket{\psi_0}\ene
on the Bloch sphere is
\bne \theta (t) = \theta_0, \quad \phi (t) = \phi_0 + \Delta E t \ene
where $\Delta E := E_0 - E_1$. The trajectory of the time evolution is shown in Fig.~\ref{fig:Bloch sphere}. Assuming that there is an underlying Nielsen geometry, then Krylov complexity equals the length of the shortest path between any pair of points on the time-evolution curve. 
This shortest path need not coincide with the curve generated by time evolution.

The Krylov complexity of $\ket{\psi(t)}$ is
\bne\label{eq:p1} \begin{split}
\cC(t) &= \sum_{n=0} n p_n (t) = p_1 (t) = \sin^2 (\theta_0) \sin^2 (\Delta E t /2)
\end{split} \ene
%This is the Krylov complexity for an arbitrary initial state. 
It is time-translation invariant because it does not depend on $\phi_0$.
It fails to satisfy the triangle inequality, as can be explicitly checked for three points at $t_A$, $t_B$, and $t_C$, and because $\sin^2(\Delta E t/2)$ is not a subadditive function of $t$. Therefore, the Krylov complexity between time-evolved states on the Bloch sphere cannot be a measure of distance between the states; there is no metric for which Krylov complexity can be reformulated as a Nielsen complexity on the Bloch sphere.

\section{General case}\label{sec:general quantum}
Next, we will show that Krylov complexity fails to satisfy the triangle inequality, using only its definition and not assuming anything about the reference state, the Hamiltonian, or the Hilbert space dimension. We will do this by Taylor expanding $\cC(t)$ and showing that it is not a subadditive function for small $t$. Note that Krylov complexity is time-reversal invariant, $\cC(t) = \cC(-t)$, as follows from the Schr\"odinger equation, so the Taylor expansion only has even powers of $t$.

We first expand the probabilities of the time-evolved state in the Krylov basis to express (\ref{eq:Complexity}) as:
\bne \begin{split} \cC(t) &= \sum n \left (p_n (0) + \frac{1}{2} \ddot{p}_n (0) t^2 + ... \right)~.
\end{split}\ene 
We'll expand to the fourth order. By definition, $p_n (0) = \delta_{n,0}$. Using the Schr\"odinger equation, the second derivatives of $p_n$ are
\bne \ddot{p}_0 (0) = -2b_1^2, \quad    \ddot{p}_1 (0) = 2b_1^2, \quad    \ddot{p}_{n>1} (0) = 0~, \ene
while the fourth derivatives are
\bne \begin{split}
p_0^{(4)}(0)&=2b_1^2((a_0-a_1)^2+4b_1^2+b_2^2),\\
    p_1^{(4)}(0)&=-2b_1^2((a_0-a_1)^2+4(b_1^2+b_2^2)),\\
    p_2^{(4)}(0)&=6b_1^2b_2^2,\quad p_{n>2}^{(4)}(0)=0.
\end{split} \ene
The Taylor expansion of the Krylov complexity is then
\begin{align}\label{eq: delta C}
    \cC(t) =b_1^2 t^2+\tfrac{b_1^2\qty(2(b_2^2-2b_1^2)-(a_0-a_1)^2)}{12} t^4+\mathcal{O}( t^6)~.
\end{align}
We immediately see that Krylov complexity will fail to satisfy the triangle inequality for three states separated by short time intervals because, for small $t$, $\cC(t) \approx b_1^2 t^2$, which is not subadditive. To be more explicit, if we choose $t_C-t_B = t_B - t_A = \delta t$, with $t_A < t_B < t_C$, then the triangle inequality is violated because
\bne \begin{split}
    2 \cC(\delta t) - \cC(2 \delta t) = - 2 b_1^2 \delta t^2 + O(\delta t^4) 
    \ngeq 0.
\end{split} \ene
While it might seem that we could avoid this conclusion if $b_1 = 0$ if this is true then $\ket{\psi_0}$ is an energy eigenstate and this is the trivial case because there is only one Krylov basis state.

We have not assumed or used anything other than definitions to show that Krylov complexity always violates the triangle inequality for three states separated by short time intervals. Thus, there is no possible metric such that Krylov complexity equals the length of the shortest path between states because the triangle inequality is a defining property of metric spaces.

\subsection{Modifications}
Lastly, for completeness, we will consider a few simple modifications to the definition of Krylov complexity, to see if any can be measures of distance in a metric space.

We could replace $n \to f(n)$ in the definition, with $f$ any monotonically increasing function. The motivation for doing so is that, for any such $f$, the Krylov basis is still the special minimising basis amongst orthonormal bases of $\sum f(n) |\bra{\psi(t)}\ket{B_n}|^2$ \cite{Balasubramanian:2022tpr}. This replacement does not help however because Taylor expansion about $t=0$ becomes $C(t) = f(1) b_1^2 t^2 + O(t^4)$ which is still not a subadditive function.

Since the triangle inequality is violated because the leading order term in~\eqref{eq: delta C} is quadratic in $t$, it is natural to try taking the square root and modifying the Krylov complexity definition to
\begin{align}\label{eq:s as K sqrt}
    \sqrt{\cC(t)}
    = b_1 t - \tfrac{((a_0-a_1)^2 + 2(2b_1^2 - b_2^2))}{24}t^3 + O(t^5)~.
\end{align}
While triangle inequality for $\sqrt{\cC(t)}$ is now saturated at leading order for small $t$, it is still violated if the sign of the cubic term is positive. The sign of the cubic term is not definite, it depends on the Lanczos coefficients, but if it can be positive, then $\sqrt{\cC(t)}$ is not a measure of distance in a metric space. 
We can give an example. For the Hamiltonian $H=\alpha(L_++L_-)$,
where $L_\pm$ are raising and lowering operators in an $sl(2,\,\mathbb{R})$ algebra, the square root of the Krylov complexity  of the time-evolved highest weight state is~\cite{Balasubramanian:2022tpr} $\sqrt{\cC(t)}\propto\sinh \alpha t.$
This is not a subadditive function of time and so does not satisfy the triangle inequality.

A third Krylov-type complexity proposal or modification that we consider is the exponential of the K-entropy~\cite{Caputa:2021sib}. This is defined as 
\bne e^{S_K (t)} := e^{-\sum_n p_n \log p_n}. \ene 
This is time-translation invariant, so we only need to check if it is a subadditive function of $t$. It is not, because its Taylor expansion is
\bne e^{S_K (t)} - e^{S_K (0)} = - (\log(b_1^2 t^2))b_1^2 t^2 + \dots \ene
where the ellipsis denotes higher powers in $t$.

We have considered three modifications of Krylov complexity, and shown that none are distances in a metric space. 

\section{Discussion}

Our initial motivation was to ask if Krylov complexity is a special case of Nielsen complexity, in the sense that there is a metric for which Krylov complexity is the length of the shortest path to the target. The hope was to determine the geometry from the lengths of geodesics.
Instead, we found that Krylov and Nielsen complexity are incompatible. We have shown that Krylov complexity is time translation invariant
%\sa{\footnote{In contrast, Nielsen complexity for states or operators does not need to be time translation invariant in general, there are particular conditions in the functionals involved in the definition of complexity that need to be satisfied in order for it to be invariant \cite{Baiguera:2023bhm}; see e.g. \cite{Ali:2019zcj} for a recent example in chaotic quantum systems. Following our argument, the triangle inequality does not reduce to the condition that the Nielsen complexity is a subadditive function of time.}} 
and so it must be a subadditive function of time to satisfy the triangle inequality and be a measure of distance between states. We have given some explicit examples where the triangle inequality between time-evolved states is violated: the two-energy level quantum system and a particle moving on the $SL(2,\mathbb{R})$ manifold, and studied the general case.

Our results do not imply that other measures of complexity between states related by time evolution need to be subadditive functions of time. This would be inconsistent with the expectation that complexity grows exponentially in time in generic chaotic systems~\cite{Gautam:2023bcm, Alishahiha:2022anw,Erdmenger:2023wjg}. 
The reduction of the triangle inequality to requiring subadditivity as a function of time occurred because of the special property that Krylov complexity is time translation invariant.
%\footnote{Nielsen complexity is also time-translation invariant. We thank Michal P. Heller for pointing this out.}
On physical grounds, we do not expect complexity to be time translation invariant; the change in complexity should depend on both the initial time and the interval of time elapsed. It may be that a coarse-grained notion of complexity is needed to capture this intuition.

Whether Nielsen complexity is a subadditive function of time depends on assumptions about the metric, and whether we are talking about operator or state complexity. Subadditivity follows from the triangle inequality and time-translation invariance.
If we assume that the metric on the space of unitary operators is right-invariant, then the Nielsen operator complexity of the time evolution operator $e^{iHt}$ for a time-independent Hamiltonian is time translation-invariant: $C(e^{iHt_1}, e^{iHt_2}) = C(\mathbbm{1},e^{iH(t_2-t_1)})$. So, with a right-invariant metric, the Nielsen complexity of the time evolution operator must be a subadditive function of time.  
%The physical justification for assuming right-invariance is that cost shouldn't depend on the position
In contrast, Nielsen state complexity does not need to be a subadditive function of time (see eq (159) in~\cite{Chapman:2018hou} for an explicit example). Since Nielsen state complexity equals Nielsen operator complexity minimised over all unitaries that map to the target state from the reference state, the unitary that minimises the Nielsen operator complexity between $\ket{\psi(t_2)}$ and $\ket{\psi(t_1)}$ will not generally be the time evolution operator $e^{iH(t_2 - t_1)}$, and the time translation invariance of the latter does not imply invariance of the former. 

Our results are not in conflict with the identification of Krylov complexity with wormhole length found in~\cite{Rabinovici:2023yex}.
%\sa{\footnote{Notice as well that a distance in the bulk theory does not necessarily represent distance for its dual, and thus Krylov does not need to satisfy the axioms of a metric space.}} 
Distance is the length of the shortest path between two points. The wormhole length is the length of a particular curve between
two points, but it is not the distance between them.

We have not ruled out Krylov complexity equalling distance between states in a more exotic geometry, such as a semimetric space. In semimetric spaces, also known as distance geometries, distances need not satisfy the triangle inequality. Nevertheless, we have ruled out Krylov complexity being a continuous version of circuit complexity, or Nielsen complexity, because those do satisfy the inequality.

Lastly, we comment on a few other interesting future directions. Our studies could be extended to identifying other complexity definitions that are not compatible with complexity being a measure of distance.
It may also be that other inequalities, such as a generalisation of the triangle inequality to volumes, could
rule out other geometric interpretations of Krylov complexity, such as volume.
Recent developments propose a notion of Krylov complexity which unifies the operator and state approaches \cite{Alishahiha:2022anw}, and 
it would be interesting to verify our results with generalizations of Krylov complexity that allow for the reference state to be a reduced density matrix of generic subregions \cite{Alishahiha:2022anw}.
For Krylov complexity for open quantum systems \cite{Liu:2022god,Bhattacharya:2022gbz,Bhattacharjee:2022lzy,Bhattacharya:2023zqt,Bhattacharjee:2023uwx}, since our work assumes unitary evolution, some of our arguments would be modified for these systems, but we do not expect that this would change the conclusions about the triangle inequality (\ref{eq: s triangle}) being violated.

\begin{acknowledgments}
We thank Vijay Balasubramanian, Souvik Banerjee, Pawel Caputa, Arghya Chattopadhyay, Ben Craps, Matt Headrick, Michal Heller, Dominik Neuenfeld, Kyriakos Papadodimas, and Dimitrios Patramanis for useful discussions. SEAG thanks the University of Amsterdam, the Delta Institute for Theoretical Physics, and the International Centre for Theoretical Physics for their hospitality and financial support during several phases of the project, and the Research Foundation - Flanders (FWO) for also providing mobility support. The work of SEAG is partially supported by the FWO Research Project G0H9318N and the inter-university project iBOF/21/084. The work of AR is supported by the Stichting Nederlandse Wetenschappelijk Onderzoek Instituten (NWO-I) through the Scanning New Horizons project, by the Uitvoeringsinstituut Werknemersverzekeringen (UWV), by FWO-Vlaanderen project G012222N, and by the Vrije Universiteit Brussel through the Strategic Research Program High-Energy Physics. 
\end{acknowledgments}

\end{document}